\documentclass[twocolumn]{aastex631}

\usepackage{xspace}
\usepackage{multirow}
\usepackage{subfigure}
\usepackage[dvipsnames, svgnames]{xcolor}
\usepackage{soul}
\usepackage[makeroom]{cancel}
\usepackage{hyperref}
\usepackage{bm}

\newcommand{\HI}{H\,\textsc{i}\xspace}
\newcommand{\hi}{H\,\textsc{i}\xspace}
\newcommand{\OIII }{[{O}\,{\footnotesize III}] \,}

\newcommand{\lrto}[1]{\textcolor{black}{#1}}
\newcommand{\lrtb}[1]{\textcolor{black}{#1}}
\newcommand{\lrtp}[1]{\textcolor{black}{#1}}
\newcommand{\lrtP}[1]{\textcolor{black}{#1}}
\newcommand{\lrtM}[1]{\textcolor{black}{#1}}
\definecolor{myLime}{HTML}{32CD32}
\newcommand{\lrtg}[1]{\textcolor{black}{#1}}
\newcommand{\lrty}[1]{\textcolor{black}{#1}}

\newcommand{\lrtnew}[1]{\textcolor{black}{{#1}}}
\newcommand{\lrtneweq}[1]{\mathcolor{black}{\bm{#1}}}

\newcommand{\lrt}[1]{#1}  

\def\Sersic{S\'{e}rsic}
\def\simba{\textsc{Simba} \,}

\begin{document}

\title{A Spatially Resolved {H}\,{\footnotesize I} Survey of Seyfert Galaxies: the Role of AGN Feedback in Shaping Atomic Gas Reservoirs}

\author{Ruitian Li}
\affil{National Astronomical Observatories, Chinese Academy of Sciences, Beijing 100101, China}
\affil{School of Astronomy and Space Science, University of Chinese Academy of Sciences (UCAS), Beijing 100049, China}

\author[0000-0002-9373-3865]{Xin Wang}
\affil{School of Astronomy and Space Science, University of Chinese Academy of Sciences (UCAS), Beijing 100049, China}
\affil{National Astronomical Observatories, Chinese Academy of Sciences, Beijing 100101, China}
\affil{Institute for Frontiers in Astronomy and Astrophysics, Beijing Normal University,  Beijing 102206, China}

\author[0000-0001-9773-7479]{Daizhong Liu}
\affil{Purple Mountain Observatory, Chinese Academy of Sciences, 10 Yuanhua Road, Nanjing 210023, China}

\author{Hui Shi}
\affil{National Astronomical Observatories, Chinese Academy of Sciences, Beijing 100101, China}
\affil{Key Laboratory of Radio Astronomy and Technology, Chinese Academy of Sciences, A20 Datun Road, Chaoyang District, Beijing 100101, China}

\author[0009-0005-3823-9302]{Yuxuan Pang}
\affiliation{School of Astronomy and Space Science, University of Chinese Academy of Sciences (UCAS), Beijing 100049, China}

\author{Pengfei Ren}
\affil{School of Astronomy and Space Science, University of Chinese Academy of Sciences (UCAS), Beijing 100049, China}

\author[0009-0007-6655-366X]{Shengzhe Wang}
\affiliation{School of Astronomy and Space Science, University of Chinese Academy of Sciences (UCAS), Beijing 100049, China}
\affiliation{National Astronomical Observatories, Chinese Academy of Sciences, Beijing 100101, China}

\author{Ming Zhu}
\affil{National Astronomical Observatories, Chinese Academy of Sciences, Beijing 100101, China}
\affil{Guizhou Radio Astronomical Observatory, Guizhou University, Guiyang 550000, China}


\author[0000-0002-5815-2387]{Mengting Ju}
\affiliation{School of Astronomy and Space Science, University of Chinese Academy of Sciences (UCAS), Beijing 100049, China}

\author[0009-0006-0596-9445]{Xiao-Lei Meng}
\affil{National Astronomical Observatories, Chinese Academy of Sciences, Beijing 100101, China}

\author[0000-0002-7020-4290]{Xinwen Shu}
\affiliation{Department of Physics, Anhui Normal University, Wuhu, Anhui 241002, China}

\author[0000-0002-2169-0472]{Ningyu Tang}
\affiliation{Department of Physics, Anhui Normal University, Wuhu, Anhui 241002, China}

\author[0000-0002-6593-8820]{Jing Wang}
\affiliation{Kavli Institute for Astronomy and Astrophysics, Peking University, Beijing 100871, China}

\author[0000-0002-4428-3183]{Chuan-Peng Zhang}
\affiliation{National Astronomical Observatories, Chinese Academy of Sciences, Beijing 100101, China}
\affiliation{Guizhou Radio Astronomical Observatory, Guizhou University, Guiyang 550000, China}

\author[0000-0003-1632-2541]{Hong-xin Zhang}
\affiliation{Department of Astronomy, University of Science and Technology of China, Hefei 230026, China}
\affiliation{CAS Key Laboratory for Research in Galaxies and Cosmology, Department of Astronomy, University of Science and Technology of China, Hefei, 230026, China}

\author{Le Zhang}
 \affil{School of Physics and Astronomy, Sun Yat-Sen University, Zhuhai 519082, China}

\author{Zheng Zheng}
\affil{National Astronomical Observatories, Chinese Academy of Sciences, Beijing 100101, China}

\author{Fujia Li}
\affil{Department of Astronomy, Tsinghua University, Beijing 100084, China}

\author[0000-0003-0062-4705]{Chen Xu}
\affil{National Astronomical Observatories, Chinese Academy of Sciences, Beijing 100101, China}
\affiliation{School of Astronomy and Space Science, University of Chinese Academy of Sciences (UCAS), Beijing 100049, China}

\author[0000-0003-4813-8482]{Sijia Li}
\affil{Department of Astronomy, Xiamen University, Xiamen, Fujian 361005, China}
\affil{School of Astronomy and Space Science, University of Chinese Academy of Sciences (UCAS), Beijing 100049, China}

\author[0000-0002-0663-814X]{Yiming Yang}
\affil{National Astronomical Observatories, Chinese Academy of Sciences, Beijing 100101, China}
\affil{School of Astronomy and Space Science, University of Chinese Academy of Sciences (UCAS), Beijing 100049, China}

\author[0009-0004-7133-9375]{Hang Zhou}
\affil{School of Astronomy and Space Science, University of Chinese Academy of Sciences (UCAS), Beijing 100049, China}

\correspondingauthor{Xin Wang}
\email{xwang@ucas.ac.cn}

\begin{abstract}

Active galactic nucleus (AGN) feedback is a key ingredient in galaxy evolution, yet its impact on the cold atomic gas reservoir --- the neutral hydrogen (\HI) phase --- remains poorly constrained. We present the most extensive spatially resolved \HI 21-cm survey of Seyfert AGN hosts to date, based on observations with the Giant Metrewave Radio Telescope (GMRT). 
\lrtnew{We measure \HI masses and sizes for eight Seyfert galaxies, and map the detailed kinematics and surface densities for two representative targets.}
We find that AGN-host galaxies exhibit a slightly shallower \HI mass–size relation than the canonical relation or the \simba simulation predictions; however, the measured slope remains consistent with the canonical value within $2\sigma$ uncertainties. This result suggests that AGN feedback does not significantly disrupt the global extent or large-scale structure of atomic gas reservoirs. To investigate the internal \HI kinematics in greater detail, we perform a 3D kinematic forward modeling of the \HI disk in UGC 4503. Our analysis reveals an elevated intrinsic velocity dispersion of $\sigma = 14.9^{+6.1}_{-3.8}$ km/s and a reduced level of rotational support, with $V/\sigma = 14.28_{-4.17}^{+4.97}$, compared to large-sample star-forming spirals. These kinematic signatures, together with localized residuals in the velocity field, indicate that AGN-driven outflows or jets may inject or indirectly affect the turbulence in the atomic gas disk, potentially regulating the cold gas reservoir. Future GMRT observations, combined with optical integral-field spectroscopy from MaNGA, will enable quantitative constraints on the role of AGN feedback in regulating star formation efficiency across a larger and more representative galaxy sample.

\end{abstract}

\keywords{Galaxies: Seyfert, Galaxies: ISM}

\section{Introduction} 
\label{sec:intro}
Feedback from active galactic nuclei (AGNs) is widely acknowledged as a fundamental mechanism in galaxy formation and evolution. Many models invoke AGN feedback as a key
ingredient for galaxy evolution \citep{springel2005modelling,hopkins2006unified,ciotti2010feedback, scannapieco2012aquila, venturi2017ionized,dave2019simba}. The energy released by supermassive black holes (BHs) at the centers of galaxies can shape the properties of the surrounding interstellar medium (ISM) through radiative, mechanical, and gravitational processes, fundamentally altering the star formation history, kinematics, and morphology of their hosts \citep{croton2006many}. AGN feedback manifests itself through various phenomena, \lrto{including} high-energy radiation, relativistic jets, and fast outflows from AGNs that interact with the ISM \citep{fabian2012observational}. 

\lrto{Numerous studies have provided analysis through observations of bubbles, PAH destruction and dust dynamics \citep{diamond2012pahlowz, fabian2012observational, cicone2014massive, yang2016agn, jensen2017pahlowz,narayanan2023pahsimu} in AGN feedback.}
\lrt{Neutral hydrogen (\HI) is regarded as a premier probe for mapping the dynamical mass distributions within galaxies \citep{de2008high, oh2015high,das2020tracing}. Its effectiveness stems not only from its properties as a cold, dissipative medium but also from its vast spatial extent, typically spanning twice the radius of the stellar disk \citep{swaters2002westerbork, bigiel2010extremely}. This makes \HI an excellent tool for tracing galaxy rotation curves, constraining dynamical mass \citep{bosma198121,van1985distribution,begeman1989hi,ho2007bulge,yu2020determination} and the dark matter distribution in galaxies \citep{katz2019tight}. }

\lrtp{So far, it remains challenging to describe the feedback from supermassive BHs on the \HI content of galaxies.} Supermassive BHs interact with the surrounding ISM, in particular \hi gas, through
high-energy radiation, ejections of matter, and gravitational interactions. High-energy rays from the BH collide with \hi gas, which heats the gas up, causing it to expand and outflow. More intense radiation causes neutral hydrogen atoms to become ions, potentially creating a region of ionized gas that affects the distribution and flow of surrounding matter. \lrtp{Recent studies have suggested that AGNs may regulate molecular gas reservoirs through heating or removal \citep[e.g.,][]{shangguanGasContentEfficiency2018,shangguanAGNFeedbackStar2020,ellison2021edge, Piotrowska2022agn}, but the role of \HI in mediating these processes remains underexplored.} \lrto{In some cases in the EAGLE simulation, radiation can even cause the decomposition of neutral hydrogen gas} \citep{crain2015eagle, schaye2015eagle, bahe2016eagle}. This means that the gas will gradually deplete and may affect the formation and evolution of galaxies. 

\lrt{By analyzing a large sample of nearby galaxies with measurements of both BH and \HI masses, \cite{wang2024black} demonstrated a scenario of how BHs regulate cool gas content in galaxies. The accumulated energy from BHs is vital in regulating the accretion
and/or cooling of cool gas in galaxies and AGN feedback ejects cool gas and
heats the circumgalactic medium (CGM), leading to low cool gas content.}

\lrto{Mapping the distribution of \hi is essential for deciphering the evolutionary pathways of galaxies \citep{Yu_2022Centrally}. Previous studies have found that there is a tight correlation between galaxies' \HI size and mass, which is known as the \hi mass–size relation \citep{broeils1997short, wang2016new}. 
Notably, \cite{wang2016new} exclusively examined nearby spiral and dwarf irregular galaxies, along with several elliptical galaxies, obtaining a well-characterized \hi mass-size relation with a scatter of approximately 0.06 dex, which revealed a uniform trend across diverse galaxy populations.}

A critical gap concerns whether AGN activity disturbs \HI surface density profiles, which would violate the classical mass-size relation. \lrto{Although \cite{wang2016new} covered a wide range of sample galaxies, they have not paid special attention to AGN host galaxies.}
If AGN indeed influences \HI surface density profiles, a pertinent question emerges: can the classical \HI mass-size relation be applied to galaxies hosting AGN? We aim to determine whether the \HI disk in AGN hosts adheres to the previous relation. 

In molecular and ionized gas phases, extensive observations have confirmed the presence of AGN-driven outflows \lrto{\citep{veilleux2020cool, zhang2023jwst, Costa2024JWST,Hermosa2024JWST,esparza2025jwstandalma}, such as CO emission extending on kpc scales \citep{cicone2014massive, Burillo2021alma, davies2024jwst} and the \OIII -- emitting cone in active galaxies \citep{venturi2017ionized, revalski2021hst,singha2022close}}. However, similar evidence in the \HI phase driven by AGNs remains rather poorly explored and the AGN feedback mechanism has not been quantitatively understood \citep{kormendy2013coevolution}. The existing galaxy samples with resolved \HI mapping are incapable of probing AGN feedback.  Spitzer Photometry \& Accurate Rotation Curves \citep[SPARC;][]{lelli2016sparc} is by far the largest sample of spiral/disk galaxies, in which only 5 of the 175 galaxies have AGN signatures according to the Sloan Digital Sky Survey \citep[SDSS;][]{york2000sloan} classification. Among the 5 AGNs, only 1 is a Seyfert galaxy. It is thus imperative to obtain  resolved \HI  observations of Seyfert AGN host galaxies \citep{ho2008new,ho2008properties,Zuo_2022}.  

\lrto{To obtain the high angular resolution and high sensitivity observations of the \HI 21 cm lines,} we conducted the largest resolved \HI survey to date of Seyfert galaxies based on the Giant Metrewave Radio Telescope (GMRT). This paper presents initial results from a carefully selected sample of 8 Seyfert galaxies (UGC 4503, UGC 6128, UGC 9535, UGC 10108, UGC 8570, UGC 8587, UGC 5103, UGC 9110), focusing on \HI distribution and kinematics – yielding measurements inaccessible to single-dish telescopes like Arecibo or the Five-hundred-meter Aperture Spherical Radio Telescope \citep[FAST,][]{nanFivehundredmeterApertureSpherical2011,jiangCommissioningProgressFAST2019}. \lrt{In combination with the kinematic models, we obtain reliable inclination angle, dynamic mass, dark matter distribution and $V/\sigma$ of UGC 4503 --- an example Seyfert galaxy in our sample to showcase the spatially resolved analysis promised by our resolved \HI data.}

The following sections are organized as follows. Section \ref{sec:Obs} details our sample selection, GMRT observations, and data reduction procedures. Section \ref{sec:res} presents the HI spectra, kinematic moment maps, measurements of HI mass and size and \HI dynamics modeling in a case study. We outline future work in section \ref{sec:future}. Throughout this paper, we adopt a standard $\Lambda$CDM cosmology with $H_0=70~\rm km \ s^{-1}\ Mpc^{-1}$, $\Omega_M=0.3$, and $\Omega_\Lambda = 0.7$.

\section{Observations and Data Reduction}
\label{sec:Obs}
\subsection{Sample Selection and GMRT Observations}

GMRT achieves high-resolution imaging through a hybrid array of 30 steerable 45 m dishes \citep{swarup1991giant}. By positioning 14 antennas in a random central pattern ($\sim$ 1 km) and spreading 16 antennas along three extended arms, the telescope creates a Y-shaped configuration with baselines spanning $\sim$ 100 m to 25 km. This arrangement optimizes $uv$ coverage and sensitivity, significantly enhancing imaging performance (highest angular resolution achievable about 2 arcsec at 1.4 GHz) \citep{rhee2018neutral}. \lrty{This layout thus enables the sensitive detection of extended \hi gas and resolves detailed kinematics.}

\lrto{We began by} cross-matching the catalogs between the seventh data release of the Sloan Digital Sky Survey 
 \citep[SDSS DR7;][]{abazajian2009seventh} and Arecibo Legacy Fast ALFA (ALFALFA; hereafter a100 in figures) survey \citep{haynes2018arecibo},
to establish conservative selection criteria, and we selected the AGN host galaxies with ALFALFA detection.

\begin{figure*}
    \centering
    \includegraphics[width=\linewidth]{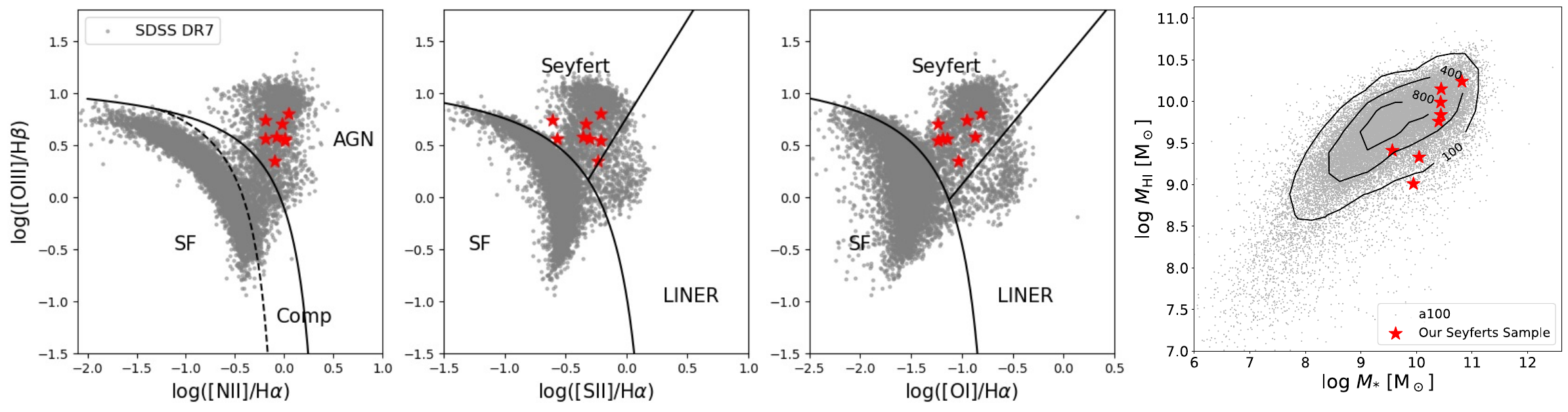}
    \caption{General properties of our Seyfert galaxies. The red stars mark our targets in this work. The three leftmost panels are BPT diagrams. The dashed curve \citep{kauffmann2003host}
     and the solid lines \citep{Kewley2006host} effectively divide all SDSS DR7 galaxies
    (dark grey dots) into star-forming (SF), Seyfert, Low Ionization Nuclear Emission-Line Region (LINER), and composite (Comp) galaxies. The right graph plots logarithm \HI mass vs. stellar mass in solar units. The light gray dots are galaxies in ALFALFA survey and 
    contours represent the 100, 400, 800 number density levels within a 0.25x0.25 bin.
    }
    \label{fig:sample}
\end{figure*}

Subsequently, we categorized the AGN types through Baldwin, Phillips, $\&$ Terlevich (BPT) diagrams \citep{baldwin1981classification, Veilleux1987} to select Seyfert galaxies, which are shown in Figure \ref{fig:sample} (left panels). \lrty{Emission-line data were retrieved from the Max-Planck-Institute for Astrophysics-John Hopkins University (MPA-JHU) database\footnote{\url{https://wwwmpa.mpa-garching.mpg.de/SDSS/}}.} In the right panel of Figure \ref{fig:sample}, we also plot logarithm \HI mass vs. stellar mass in solar units and number density contours, which demonstrate that our GMRT observations can secure the homogeneous sample of local Seyfert galaxies. \lrtnew{
Since ALFALFA provides only total \HI masses, we used the \cite{wang2016new} relation to select Seyfert galaxies with initial \HI diameter estimations (\HI size $>2$ arcmin, surface density at 1 $\rm M_{\odot}/pc^{2}$ level). This criterion ensures that the disks are sufficiently resolved for kinematic analysis. The \HI diameters reported in this paper are measured directly and independently from our GMRT data.}
Finally, we restricted the signal-to-noise ratio (SNR) of \HI spectrum to be over 20. Basic properties of our 8 carefully selected Seyfert AGN host galaxies are listed in Table \ref{tab:info}.

\begin{deluxetable*}
{ccccccc}   
\tablecolumns{7}
\tabletypesize{\scriptsize}
\tablewidth{0pt} 
\tablecaption{Basic information of our Seyfert galaxy sample.}
\tablehead{
\colhead{Galaxy} & 
\colhead{RA [J2000]}&  
\colhead{DEC [J2000]} &
\colhead{$d$ [Mpc]}&
\colhead{log $M_{*}$ [M$_\odot$]}&
\colhead{SNR}&
\colhead{GMRT Observation Date}\\
\colhead{(1)} & 
\colhead{(2)} & 
\colhead{(3)} & 
\colhead{(4)} & 
\colhead{(5)} & 
\colhead{(6)} & 
\colhead{(7)} 
} 

\startdata 
UGC 4503$^{a}$  & $13^\mathrm{h}\,35^\mathrm{m}\,19^\mathrm{s}.24$&
$+26^\mathrm{d}\,25^\mathrm{m}\,29^\mathrm{s}.10$ & 62.2  & 10.82 & 90.1&2024/08/24  \\
UGC 6128$^{a}$  & $11^\mathrm{h}\,04^\mathrm{m}\,02^\mathrm{s}.95$ & $+28^\mathrm{d}\,02^\mathrm{m}\,12^\mathrm{s}.30$ & 24.1  & 9.94  & 81.9&2024/08/25   \\
UGC 9535$^{a}$  & $14^\mathrm{h}\,48^\mathrm{m}\,42^\mathrm{s}.56$ & $+12^\mathrm{d}\,27^\mathrm{m}\,25^\mathrm{s}.90$ & 29.3  & 9.57  & 81.6&2024/08/25   \\
UGC 10108$^{a}$ & $15^\mathrm{h}\,58^\mathrm{m}\,07^\mathrm{s}.97$ & $+12^\mathrm{d}\,04^\mathrm{m}\,13^\mathrm{s}.00$ & 70.2  & 10.40 & 35.0&2024/08/30   \\
UGC 8570$^{a}$  & $13^\mathrm{h}\,35^\mathrm{m}\,19^\mathrm{s}.24$ & $+26^\mathrm{d}\,25^\mathrm{m}\,29^\mathrm{s}.10$ & 112.2 & 10.44 & 24.4&2024/08/31   \\
UGC 8587$^{a}$  & $13^\mathrm{h}\,35^\mathrm{m}\,55^\mathrm{s}.20$ & $+35^\mathrm{d}\,35^\mathrm{m}\,17^\mathrm{s}.70$ & 43.5  & 10.05 & 22.8&2024/09/03   \\
UGC 5103$^{b}$  & $09^\mathrm{h}\,34^\mathrm{m}\,57^\mathrm{s}.59$ & $+21^\mathrm{d}\,42^\mathrm{m}\,18^\mathrm{s}.90$ & 55.6  & 10.44 & 49.6&2024/11/30  \\
UGC 9110$^{b}$  & $14^\mathrm{h}\,14^\mathrm{m}\,13^\mathrm{s}.39$ & $+15^\mathrm{d}\,37^\mathrm{m}\,20^\mathrm{s}.60$ & 68.4  & 10.43 & 46.1&2024/12/02  \\
\enddata
\tablecomments{\lrto{Column (1): Galaxy ID; (2) and 3: Equatorial coordinates in J2000 respectively; (4): Adopted distance; (5): Logarithm of the stellar mass in solar units; (6): Signal-to-noise ratio (SNR).} Column (1) to (6) are obtained from ALFALFA catalogue \citep{haynes2018arecibo}. $^a$PID: 46\_116 (PI: Pengfei Ren); $^b$PID: 47\_011 (PI: Pengfei Ren); }

\end{deluxetable*}\label{tab:info}

We conducted a total of 40 hours of GMRT observations targeting 8 Seyfert
galaxies. These observations were conducted in two GMRT cycles and were carried out from August to December 2024 (Program IDs 46\_116 and 47\_011, PI: Pengfei Ren). The redshifted \HI lines of our targets have frequencies within 1366 MHz and 1414 MHz. They are within the L-band receiver frequency range (Band-5: 950-1500 MHz) and a bandwidth of 100 MHz was employed. This entire bandwidth was further divided into 8192 channels, which resulted in a frequency resolution of 12.207 kHz. We aimed for a beam size of $\sim 15''$ (at least 8 points for kinematics models \citep{kormendy2013coevolution}) to produce high-quality velocity field maps. The observed standard flux density calibrators (\lrty{3C138, 3C286, and 3C48}) were observed in the first and last scan to correct for variations in amplitude and bandpass. These scans (each lasting about
10 minutes) were used for flux density, delay, and bandpass calibrations.
In one loop, a compact radio source was utilized as a phase calibrator, which was typically scanned for 5 minutes before observations
of the target galaxy and the target source was observed for about 42 minutes. There were 5 loops in one day observation run.

\subsection{Data Reduction}
The observed GMRT data were processed by Common Astronomy Software Applications package \citep[CASA\footnote{\url{https://casa.nrao.edu/}};][]{mcmullin2007casa} and visualized by Cube Analysis and Rendering Tool for Astronomy \citep[CARTA\footnote{\url{https://cartavis.org/}};][]{ott2020carta} and we mainly followed data reduction procedures in CASA {guidelines}\footnote{\url{https://casaguides.nrao.edu/index.php/HI_21cm_(1.4_GHz)_spectral_line_data_reduction:_LEDA_44055-CASA6.5.2}}.

Initially, we inspected the data and flagged the obvious radio frequency interference 
(RFI) using ``manual'' mode. The data were flagged further and calibrated with the aid of GMRT-pipeline--CAPTURE-CASA6 \citep{kale2021capture}.
In the calibration process, we first conducted delay and bandpass calibration using the bandpass/flux calibrators and calibrated the phase and amplitude of standard flux calibrators and phase calibrators. Then, the flux density scale from the flux calibrator was transferred to the phase calibrator. After that, we inspected the results after an initial calibration, executed ``manual'' mode flagging again and re-calibrated the data to get better results. \lrto{ We performed the above mentioned flagging and calibration processes iteratively,} until we eventually obtained the satisfying performance. Lastly, final calibration tables were applied to both calibrators and the target source.

We enumerate corrected flux densities of our calibrators in Table~\ref{tab:flux}.

\begin{deluxetable*}
{ccccccc}   
\tablecolumns{7}
\tabletypesize{\scriptsize}
\tablewidth{0pt} 
\tablecaption{Flux density of the calibrator\label{tab:flux density}}
\tablehead{
\colhead{Calibrator}& 
\colhead{Type}&  
\colhead{RA [J2000]}&
\colhead{DEC [J2000]}&
\colhead{Reference Flux [Jy]}& 
\colhead{Observed Flux [Jy]}&
\colhead{Target Source}
\\
\colhead{(1)} & 
\colhead{(2)} & 
\colhead{(3)} & 
\colhead{(4)} & 
\colhead{(5)} & 
\colhead{(6)} &
\colhead{(7)}
} 

\startdata 
{3C138}& \multirow{2}{*}{Bandpass/Flux }&  \multirow{2}{*}{$05^\mathrm{h}\,21^\mathrm{m}\,09^\mathrm{s}.89$}  & \multirow{2}{*}{{$+16^\mathrm{d}\,38^\mathrm{m}\,22^\mathrm{s}.05$}}  & \multirow{2}{*}{8.47}&   8.56     &UGC 4503 \\
{3C138}& &    &   & &   8.61     &UGC 5103 \\
\noalign{\smallskip}\hline\noalign{\smallskip}
{3C286}&\multirow{6}{*}{Bandpass/Flux }& \multirow{6}{*}{$13^\mathrm{h}\,31^\mathrm{m}\,08^\mathrm{s}.29$}  &\multirow{6}{*}{$+30^\mathrm{d}\,30^\mathrm{m}\,32^\mathrm{s}.96$ }& \multirow{6}{*}{15.00} & 15.10&UGC 6128\\
{3C286}& & & & &15.11&UGC 9535\\
{3C286}& &   & & & 15.16&UGC 10108\\
{3C286}& &  & & & 15.31&UGC 8570\\
{3C286}& &  &  & & 15.35&UGC 8587\\
{3C286}& &   & & & 15.15&UGC 9110\\
\noalign{\smallskip}\hline\noalign{\smallskip}
{0854+201       }&\multirow{5}{*}{Phase }&     $08^\mathrm{h}\,54^\mathrm{m}\,48^\mathrm{s}.87$& $+20^\mathrm{d}\,06^\mathrm{m}\,30^\mathrm{s}.64$ & -- & 2.00 &UGC 4503\\
1125+261&{}&$11^\mathrm{h}\,25^\mathrm{m}\,53^\mathrm{s}.71$&  $+26^\mathrm{d}\,10^\mathrm{m}\,19^\mathrm{s}.98$&0.77&{0.73} &UGC 6128\\
{1640+123}&{} &$16^\mathrm{h}\,09^\mathrm{m}\,13^\mathrm{s}.32$  &     $+26^\mathrm{d}\,41^\mathrm{m}\,29^\mathrm{s}.04$ &2.10& 2.03   &UGC 10108\\
0842+185 &{} & $08^\mathrm{h}\,42^\mathrm{m}\,09^\mathrm{s}.46$ & $+18^\mathrm{d}\,35^\mathrm{m}\,40^\mathrm{s}.99$ &1.04 & 1.07 &UGC 5103\\
1445+099 &{} & $14^\mathrm{h}\,45^\mathrm{m}\,16^\mathrm{s}.46$ & $+09^\mathrm{d}\,58^\mathrm{m}\,36^\mathrm{s}.07$ &2.60 & 2.19 &UGC 9535\\
1445+099 &{} &$14^\mathrm{h}\,45^\mathrm{m}\,16^\mathrm{s}.46$ & $+09^\mathrm{d}\,58^\mathrm{m}\,36^\mathrm{s}.07$ &2.60 & 2.27 &UGC 9110\\
{1131+305}&{ }&$13^\mathrm{h}\,31^\mathrm{m}\,08^{\mathrm{s}}{.}29$  &$+30^\mathrm{d}\,30^\mathrm{m}\,32^\mathrm{s}.96$ & 15.00 & 15.00&UGC 8570\\
{1131+305}&{ }&$13^\mathrm{h}\,31^\mathrm{m}\,08^{\mathrm{s}}{.}29$  &$+30^\mathrm{d}\,30^\mathrm{m}\,32^\mathrm{s}.96$ & 15.00 & 15.14&UGC 8587\\
\enddata
\tablecomments{The reference flux densities and coordinates are available at  \url{https://science.nrao.edu/facilities/vla/observing/callist} and NASA/IPAC Extragalactic Database. Flux densities can be obtained in imaging processes of the calibrator and the reference flux of 0854+201 (a blazar) is omitted owing to source variability.}

\end{deluxetable*}\label{tab:flux}

After the calibration mentioned above, we split our target source from the corrected data and plot the data set averaging over all baselines and scans to identify what channels the \HI spectral line might occupy, as well as serve as a validation of the calibration accuracy. For the spectral line images, we employed the ``uvcontsub\_old'' task to conduct continuum fitting and subtraction in the uv plane. \lrto{In the imaging process}, we took ``tclean'' task in the continuum subtracted emission file and chose ``cube'' mode to perform the radio interferometric image reconstruction with natural weighting in order to have high sensitivity to the extended gas \citep{wang2016new}. 

\lrto{In detail, we started the imaging procedures} by interactively cleaning channels with masks around the expected emission and this initial cube was imported into CARTA. Then, we created a rectangular region and used the statistics widget to record the root mean square level (RMS). Based on this value, we could determine the cleaning threshold to use for our final image cube.  Eventually, we cleaned spectral channels non-interactively with masks file produced above, a threshold stopping criterion of approximately 3$\sigma$ of the noise levels \lrtnew{(as recommended in CASA guidelines)}
and an arbitrarily large maximum number of iterations to ensure the threshold was reached. 

We visualized our \HI lines and produced intensity (moment 0) maps, velocity field (moment 1) maps and velocity dispersion (moment 2) maps respectively in CARTA. We excluded values lower than 3 RMS to remove noise signals when generating moment maps and calculating the \HI integrated flux density.

\section{results}
\label{sec:res}
\lrto{In this section, we present the detailed analysis of the GMRT \HI observations for our sample of Seyfert galaxies. The results are organized as follows: Section \ref{subsec:lines_maps} describes the \HI spectral characteristics and the spatial gas distribution revealed by moment maps. In Section \ref{subsec:msr}, we examine the global scaling laws by deriving the \HI mass-size relation and comparing it with the relation in \cite{wang2016new}. Finally, Section \ref{subsec:model} presents a spatially resolved dynamical modeling case study of UGC 4503 to investigate its kinematic structure and potential signatures of AGN feedback.}

\subsection{\HI Spectral Lines and Moment Maps}
\label{subsec:lines_maps}
\begin{figure*}
    \centering
    
        \includegraphics[width=0.8\linewidth]{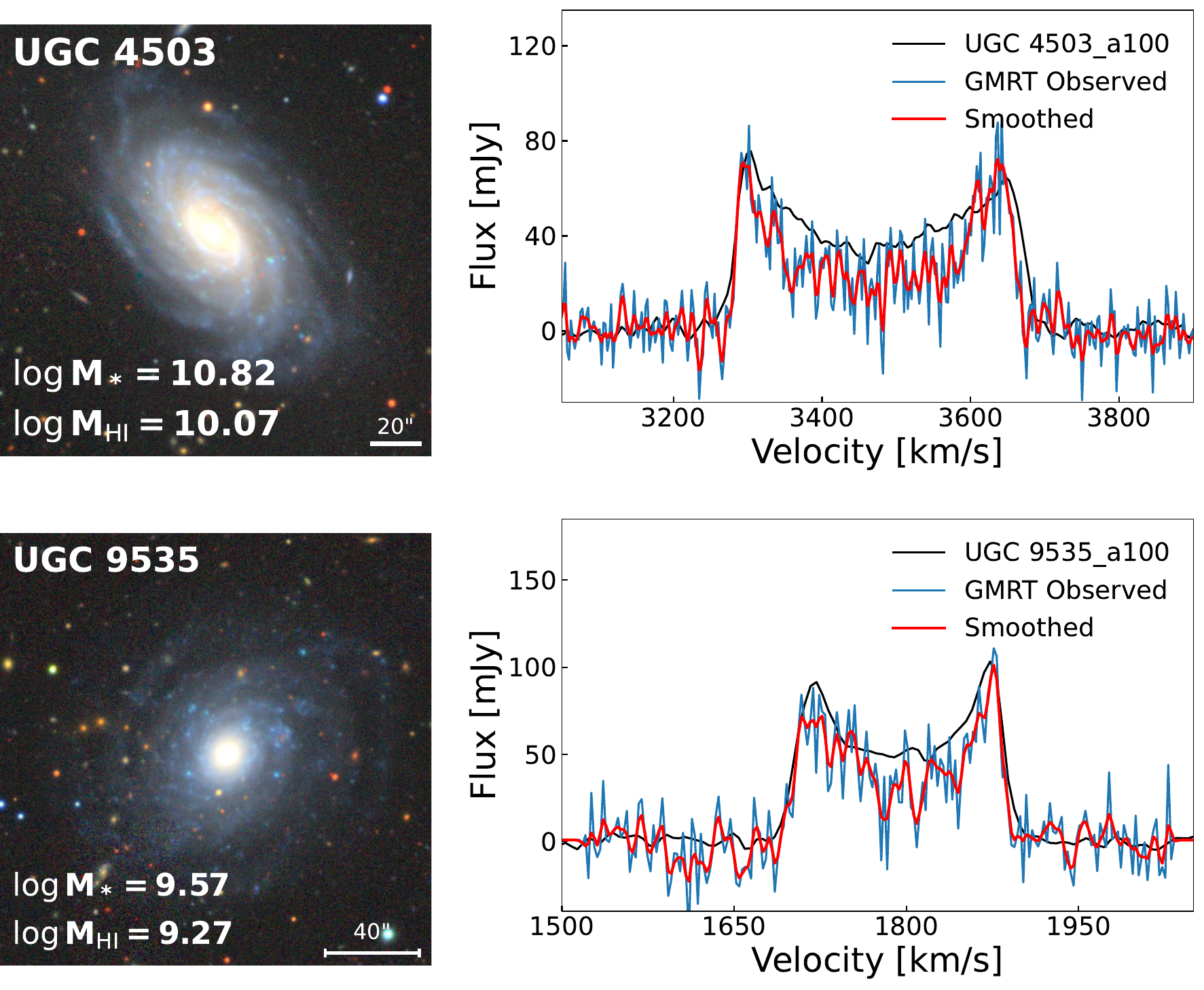}

    \caption{Optical images and \HI spectra of UGC 4503 and UGC 9535, two Seyfert galaxies in our sample. The left panels
show their DECaLS grz color-composite images \citep{dey2019overview}. In the right panels, we display their \HI spectra. The black lines illustrate ALFALFA 21 cm single dish spectra \citep{haynes2018arecibo} and blue lines correspond to our GMRT results.  \lrto{Red lines represent smoothed GMRT spectra using a Gaussian kernel with a standard deviation of $\sigma_{\mathrm{smooth}} = 1$ channel (corresponding to $\sim 2.6$ km/s). The stellar mass from ALFALFA and \HI mass from GMRT observations are written on the RGB images.}}
    \label{fig:HIspectrum}
\end{figure*}

We detected significant \lrty{\HI line} emission in all 8 Seyfert galaxies from GMRT observations after data reduction and imaging processes (see Table \ref{tab:galaxy}). Figure \ref{fig:HIspectrum} displays two representative Seyfert targets (UGC 4503 and UGC 9535) with their color-composite images from the Dark Energy Camera Legacy Survey (DECaLS; \citealt{dey2019overview}) and \HI spectral lines from ALFALFA \citep{haynes2018arecibo} and our GMRT survey. \HI emission profiles from the two telescopes exhibit the characteristic double-peaked feature, which is the signature of ordered rotation. Due to the lack of short baselines inherent to radio interferometry, the integrated \HI flux density recovered by GMRT is lower than that measured by the single-dish ALFALFA survey.

\begin{figure*}
    \centering
    \includegraphics[width=0.7\linewidth]{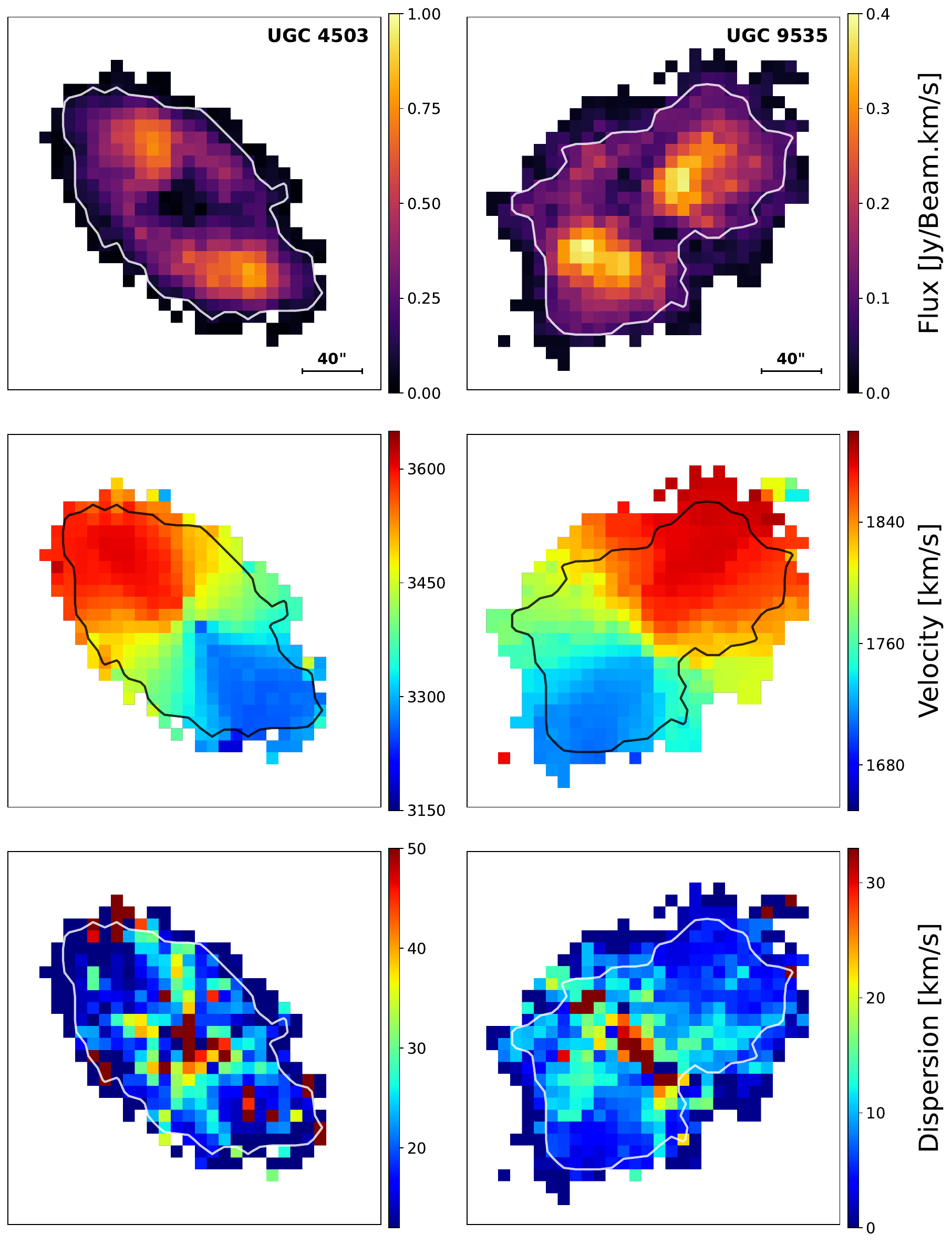}
    \caption{\lrto{Overview of \HI kinematic maps for UGC 4503 and UGC 9535. The left panels correspond to UGC 4503 and the left ones display UGC 9535}. The \HI intensity maps, velocity field maps, and velocity dispersion maps are presented (from top to bottom) respectively. The white and black contours mark \HI spatial distribution boundaries of our targets, which are defined when \HI surface densities reach 1 $\rm M_\odot/pc^{2}$ \citep{wang2016new}.
    }
    \label{fig:moments}
\end{figure*}

We also present \HI moment 0, 1, 2 maps of UGC 4503 and UGC 9535 in Figure \ref{fig:moments} as examples. The \HI intensity maps can reveal the morphology in the disk gas distribution, revealing spiral, ringed and centrally concentrated geometries \citep{haan2007atomic}. In our sample, \hi disk morphologies are visually classified. \HI morphologies of our Seyfert sample exhibit characteristic ring-like features.

\subsection{\HI Mass-Size Relation for AGN-host Galaxies}
\label{subsec:msr}
\HI integrated flux density $S_{\rm HI}$, in Jy km/s, is calculated by integrating the \HI spectral lines in the \HI emission portions, excluding flux values lower than 3 RMS. The uncertainty of the integrated flux density $\sigma_{S_{\rm HI}}$ can be estimated as the product of RMS, spectral channel numbers and velocity resolution. With integrated flux densities, we derive the logarithm of \HI mass in solar units $\log M_{\rm HI}$ from the standard formula \citep{haynes2018arecibo}, where distances $d$ are given in Table \ref{tab:info} column 4.

The uncertainty of \HI mass $\sigma_{\log M_{\rm HI}}$ is computed mainly following the methodology of \cite{jones2018alfalfa}, including the uncertainty in the integrated \HI
line flux \lrto{with a lower limit of 10\% uncertainty}. The latter minimum serves two purposes. One is to avoid unrealistically small error estimates, and the other is to account for systematic uncertainties inherent in flux calibration procedures. Here we do not consider errors in the distance, where we regard the distance of local Seyfert galaxies as known values, and therefore the equation is given as follows. 

\begin{equation}
	\sigma_{\log  M_{\mathrm{HI}}} = \frac{\sqrt{(\sigma_{S_{\rm HI}}/S_{\rm HI})^2+0.1^2}}{\ln 10}
\end{equation}

We directly measure the \HI size $D_{\mathrm{HI}}$ following the procedure of \cite{wang2014observational} based on \HI intensity maps. In the first step, $D_{\mathrm{HI,0}}$ is defined as the major axis of a fitted ellipse
to the \HI distribution where the \HI surface density reaches 1 $\rm M_\odot/pc^{2}$. We utilize LMFIT \citep{newville2016lmfit} to perform ellipse fitting. In a straightforward view, errors in the \HI size mainly arise from uncertainties in fitting.  The \HI mass can be converted into \HI surface density $\Sigma_\mathrm{HI}$ using Equation (\ref{eq:HIsurface}), where we suppose that it is a Gaussian beam with the major axis $\theta_{maj}$ and the minor axis $\theta_{min}$. 

\begin{equation}
     \Sigma_\mathrm{HI} = \frac{2.356 \times 10^5 S_{\rm {HI}}}{1.13\theta_{maj}\times\theta_{min}} 
     \label{eq:HIsurface}
\end{equation}

In the final step, the \HI diameter should be corrected for beam smearing effects
based on a Gaussian approximation \citep{wang2016new} in Equation (\ref{eq:diskcorrection}),

\begin{equation}
     \lrtneweq{D_{\mathrm{HI}} = \sqrt{D_{\rm {HI,0}}^2-\theta_{maj}\times\theta_{min}}}
     \label{eq:diskcorrection}
\end{equation}
where $D_{\mathrm{HI}}$ and $D_{\mathrm{HI,0}}$ are the corrected and uncorrected \HI sizes.
The results of the above measurements are summarized in Table \ref{tab:galaxy}.

\begin{deluxetable*}
{lccccc}   
\tablecolumns{5}
\tabletypesize{\scriptsize}
\tablewidth{0pt} 
\tablecaption{Derived {H}\,{\footnotesize I} properties of our Seyfert galaxies from GMRT.}
\tablehead{
\colhead{Galaxy} & 
\colhead{$S_{\rm HI}$ [Jy km/s]}&  
\colhead{log $M_{\rm HI}$ [M$_\odot$]} &
\colhead{$D_{\rm HI}$ [Kpc]}&
\colhead{RMS [mJy/beam]}\\
\colhead{(1)} & 
\colhead{(2)} & 
\colhead{(3)} & 
\colhead{(4)} & 
\colhead{(5)} 
} 

\startdata 
{UGC 4503}   & 13.002 & 10.074 $\pm ~0.045$ & 63.688 $\pm ~0.795$  & 1.89 \\
{UGC 6128}  & 5.380  & 8.867 $\pm ~0.044$  & 14.617 $\pm ~0.412$ & 2.53  \\
{UGC 9535}  & 9.260  & 9.273  $\pm ~0.037$ & 28.384 $\pm ~0.642$ & 2.25  \\
{UGC 10108} & 3.286 & 9.582 $\pm ~0.045$ & 31.170 $\pm ~1.606$  & 2.02                                                                \\
{UGC 8570}  & 3.437  & 10.007 $\pm ~0.044$ & 50.610 $\pm ~3.169$ & 1.55 \\
{UGC 8587}  & 3.392 & 9.179 $\pm ~0.046$  & 26.683 $\pm ~1.106$ & 3.12\\
{UGC 5103}  & 7.000 & 9.887 $\pm ~0.045$ & 36.916 $\pm ~2.062$ & 2.43 \\
{UGC 9110}  & 9.730 & 9.850 $\pm ~0.044$ & 43.011 $\pm ~2.027$ & 1.96                                                               
\enddata
\tablecomments{\lrto{Column (1): Galaxy ID; (2): \HI
line flux density of the source; (3): Logarithm of {H}\,{\footnotesize I} mass in solar units; (4): \HI diameter in kpc; (5): Root-Mean-Square (RMS) measured over the signal- and RFI-free parts of the \HI spectrum.} Flux densities of each galaxy are obtained by integrating \lrty{\HI lines} within the emission range. \HI mass can be computed via the standard formula $M_{\rm HI}=2 .356\times 10^{5}d^{2}S_{\rm HI}$\citep{haynes2018arecibo}
 using the distance given in Table \ref{tab:info}. In calculations of the \HI diameter, we also adopt distances mentioned above. }

\end{deluxetable*}\label{tab:galaxy}

\begin{figure*}
    \centering
    \includegraphics[width=0.8\linewidth]{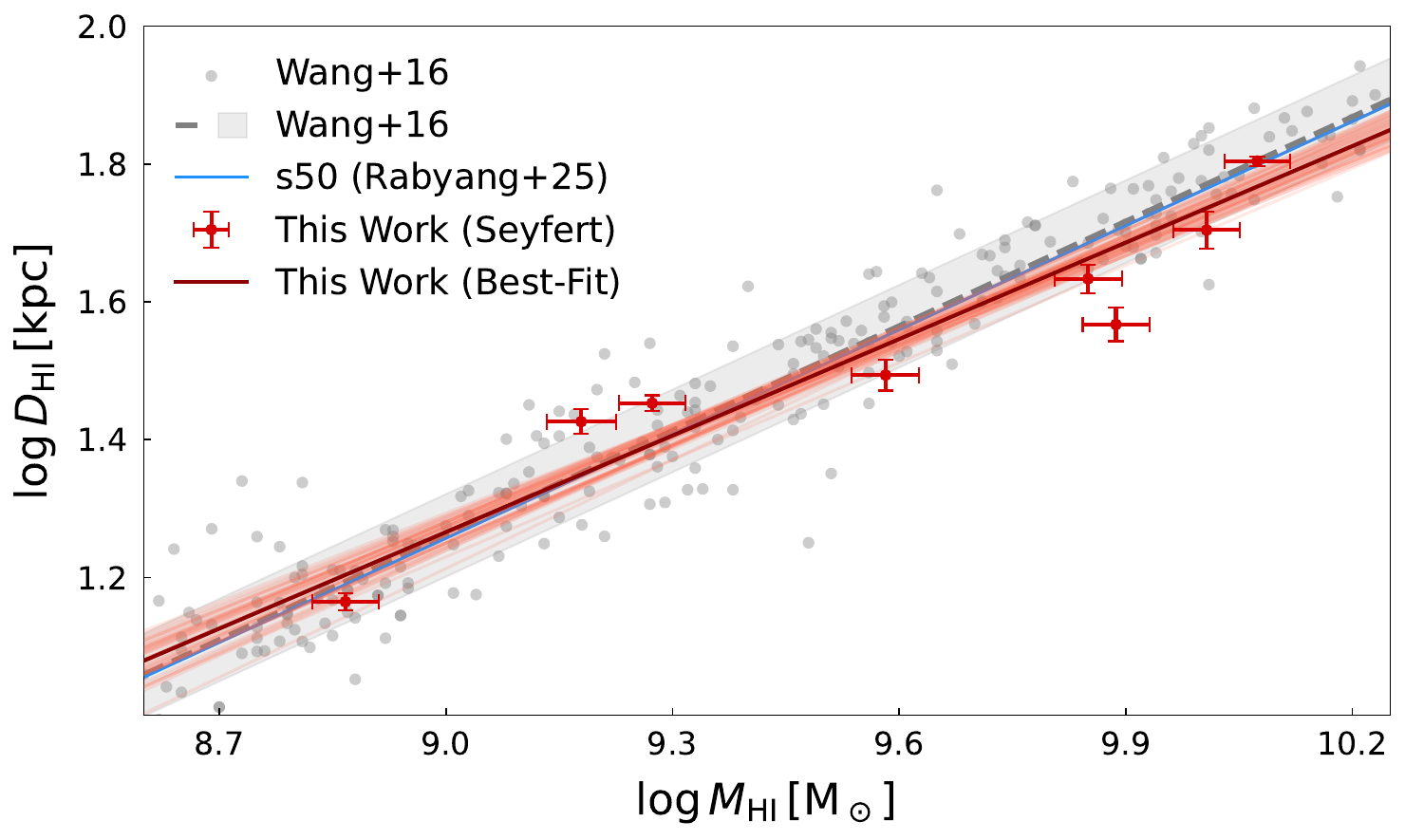}
    \caption{The \HI mass-size relation for 8 Seyfert galaxies from GMRT observations. The red points with error bars represent our Seyfert sample. \lrtg{The solid red line denotes the best-fit linear regression derived via emcee, which yields a reduced chi-squared of $\chi^2_\nu \approx 1.16$.} The thin light red lines represent 50 random draws from the fitting. \lrto{Data and the \HI mass-size relation in \cite{wang2016new} are displayed by gray marks with a shaded band of 0.06 dex \citep{wang2016new}. The blue line is the \simba simulation result (see Table \ref{tab:simba}).}}
    \label{fig:mass_size}
\end{figure*}

\lrto{After obtaining the \HI parameters and their uncertainties}, we present the \HI mass-size relation of Seyfert galaxies. 
\lrto{As shown by the red points in Figure \ref{fig:mass_size}, our Seyfert sample spans approximately 1.2 dex in $\log M_{\rm {HI}}$ and 0.6 dex in $\log D_{\rm {HI}}$. This coverage overlap demonstrates that our targets are representative.}
We perform a robust linear fit by emcee regression \citep{foreman2013emcee} to the data points accounting for measurement uncertainties in \HI mass and size mentioned above and obtain the \HI mass-size relation shown in Equation (\ref{eq:mass-size-gmrt}),
\begin{equation}
	\log D_{\rm {HI}} = (0.466 \pm 0.022) \log M_{\rm {HI}} -(2.921 \pm 0.207)
	\label{eq:mass-size-gmrt}
\end{equation}
\lrtg{with a reduced chi-squared of $\chi^2_\nu \approx 1.16$. This indicates that the linear model provides a statistically adequate description of the sample.}
The \hi mass-size relation in our work conforms to the classical relation within uncertainties in \cite{wang2016new} including dwarf galaxies, spiral galaxies and early-type disc galaxies and we demonstrate this consistency in Figure \ref{fig:mass_size}. In addition to the best-fit slope, 50 random draws from the emcee are shown, mostly within the gray shaded band (0.06 dex) of the \cite{wang2016new} relation. Based on the GMRT data for our 8 Seyfert hosts, we suggest that the outer \HI disks appear insensitive to the central AGN activity and remain consistent with the classical \HI mass-size relation.

\begin{deluxetable*}{lccccccc}
\tabletypesize{\scriptsize}
\tablewidth{0pt}
\tablecaption{Comparison of the \hi\ mass-size relation parameters from our GMRT results, Wang+2016, and \simba simulations.}
\tablehead{
    \colhead{} &
    \colhead{This Work} &
    \colhead{Wang+2016} &
    \colhead{s50$^{a}$} &
    \colhead{s50nox$^{b}$} &
    \colhead{s50nojet$^{c}$} &
    \colhead{s50noagn$^{d}$} &
    \colhead{s50nofb$^{e}$}\\
\colhead{} & 
\colhead{(1)} & 
\colhead{(2)} & 
\colhead{(3)} & 
\colhead{(4)} & 
\colhead{(5)} & 
\colhead{(6)} & 
\colhead{(7)}
}
\startdata
Sample Size          & 8                  & 562                & 904                & 1171               & 1301               & 896                & 6481               \\
$\alpha$        & $0.466 \pm 0.022$  & $0.506 \pm 0.003$  & $0.504 \pm 0.010$  & $0.530 \pm 0.008$  & $0.522 \pm 0.006$  & $0.506 \pm 0.009$  & $0.531 \pm 0.006$  \\
$\beta$         & $-2.921 \pm 0.207$ & $-3.293 \pm 0.009$ & $-3.279 \pm 0.093$ & $-3.541 \pm 0.075$ & $-3.458 \pm 0.059$ & $-3.294 \pm 0.085$ & $-3.507 \pm 0.056$ \\
$\sigma$ [dex]  &      0.064              & 0.06               & 0.096              & 0.103              & 0.083              & 0.097              & 0.122
\enddata
\tablecomments{The simulation outputs (s50 to s50nofb) are from \cite{rabyang2025simba}.
Notes on the superscripts:
$^{a}$Including stellar feedback, AGN winds, AGN jets, and X-ray heating;
$^{b}$excluding X-ray heating;
$^{c}$including stellar feedback and AGN winds only;
$^{d}$including only stellar feedback (no AGN feedback);
$^{e}$no feedback mechanisms. 
The \hi\ mass-size relation is described as $\log D_{\rm HI} = \alpha \log M_{\rm HI} + \beta$.
Row (1): Number of galaxies in the sample; 
(2): Slope ($\alpha$) of the relation; 
(3): Intercept ($\beta$) of the relation;
(4): Scatter ($\sigma$) in dex.
}
\label{tab:simba}
\end{deluxetable*}\label{tab:simba}

\lrtb{Table \ref{tab:simba} presents a comparison of the \hi mass-size relation parameters from our GMRT observations, the classical relation \citep{wang2016new}, and the recent \simba cosmological simulations \citep{rabyang2025simba}. Our derived slope ($\alpha=0.466\pm0.022$) is slightly shallower than both the classical relation ($\alpha=0.506\pm0.003$) and the simulation predictions (``s50'', $\alpha\approx0.504$). Nevertheless, the measured slope remains consistent with these canonical values within the $2\sigma$ uncertainties. This indicates that AGN feedback does not strongly disrupt the global extent or large-scale structure of atomic gas reservoirs.} Additionally, the simulation results demonstrate that the \hi mass-size relation is robust against changes in feedback mechanisms (see Table \ref{tab:simba}). These align with the analytical work by \cite{stevens2019origin}, which suggests that simulations fail to predict the \hi mass-size relation when the structure of \hi discs is disrupted through powerful feedback.

\lrt{Our \hi mass-size relation can also provide a spatially resolved perspective that complements recent global statistical studies. \cite{wang2024black} mention the role of cumulative BH accretion energy in regulating the total cold gas reservoir. In this context, our finding implies that while AGN feedback acts to regulate the \hi mass over evolutionary timescales, the host galaxies still maintain the mass-size relation in the outer disks.}

\lrtp{Due to the small sample size, current results can be  influenced by extreme outliers. Our further acquisition of resolved \HI data in a broader sample of Seyfert galaxies is critical for deriving a statistical correlation regarding AGN feedback.}

\subsection{Dynamical Modeling of \HI Maps: A Case Study of UGC 4503}
\label{subsec:model}
In this section, we investigate the connection between AGN feedback and galaxy disk dynamics of UGC 4503. 
\lrtM{This galaxy is selected as the representative case because it exhibits the highest signal-to-noise ratio and optimal data quality.}
The basic information and observed \HI results of this Seyfert galaxy are displayed in Table \ref{tab:info}, Table \ref{tab:galaxy}, Figure \ref{fig:HIspectrum} and Figure \ref{fig:moments}.
We carry out the forward dynamical modeling and MCMC-based
kinematic fitting to the \HI datacube using the
{\sc Dysmal}/{\sc DysmalPy} software \lrtM{\citep{2004ApJ...602..148D,2004ApJ...613..781D,2009ApJ...697..115C,2011ApJ...741...69D,2016ApJ...831..149W,2017ApJ...840...92L,2021ApJ...922..143P,2025ApJ...978...14L}}. {\sc Dysmal}/{\sc DysmalPy} is a physically-motivated, multi-component, three dimensions galaxy dynamical forward-modeling tool. {\color{black} From {\sc Dysmal}/{\sc DysmalPy}, we can derive an intrinsic three-dimensional dynamical model that incorporates the mass distributions of both baryonic and dark matter components, considering projection effects, beam smearing, spectral resolution, and sampling effects.} The fitting procedures, including three dimensions (fitting a data cube), two dimensions (fitting velocity and velocity dispersion maps), or one-dimensional (fitting one dimension profiles extracted from a pseudo-slit) are all conducted directly in the ``data space'' using the same extraction method as applied to the observational data.

\textsc{DysmalPy} builds up a galaxy using several physically-motivated and geometry parameters \citep{liu2023600}, e.g., bulge and disk components in \Sersic{} profiles and the dark matter halo in Navarro-Frenk-White (NFW; \citealt{NFW}) profile. 
The key unfixed parameters we use in the MCMC fitting are the total mass of disk+bulge ($M_{\rm {total}}$), the effective radius of disk in kpc ($R_{\mathrm{eff,\,disk}}$), the halo virial mass ($M_{\rm {vir, halo}}$),  the intrinsic dispersion value ($\sigma_0$), inclination, and position angle. 

\begin{figure*}
    \centering
    \includegraphics[width=0.9\linewidth]{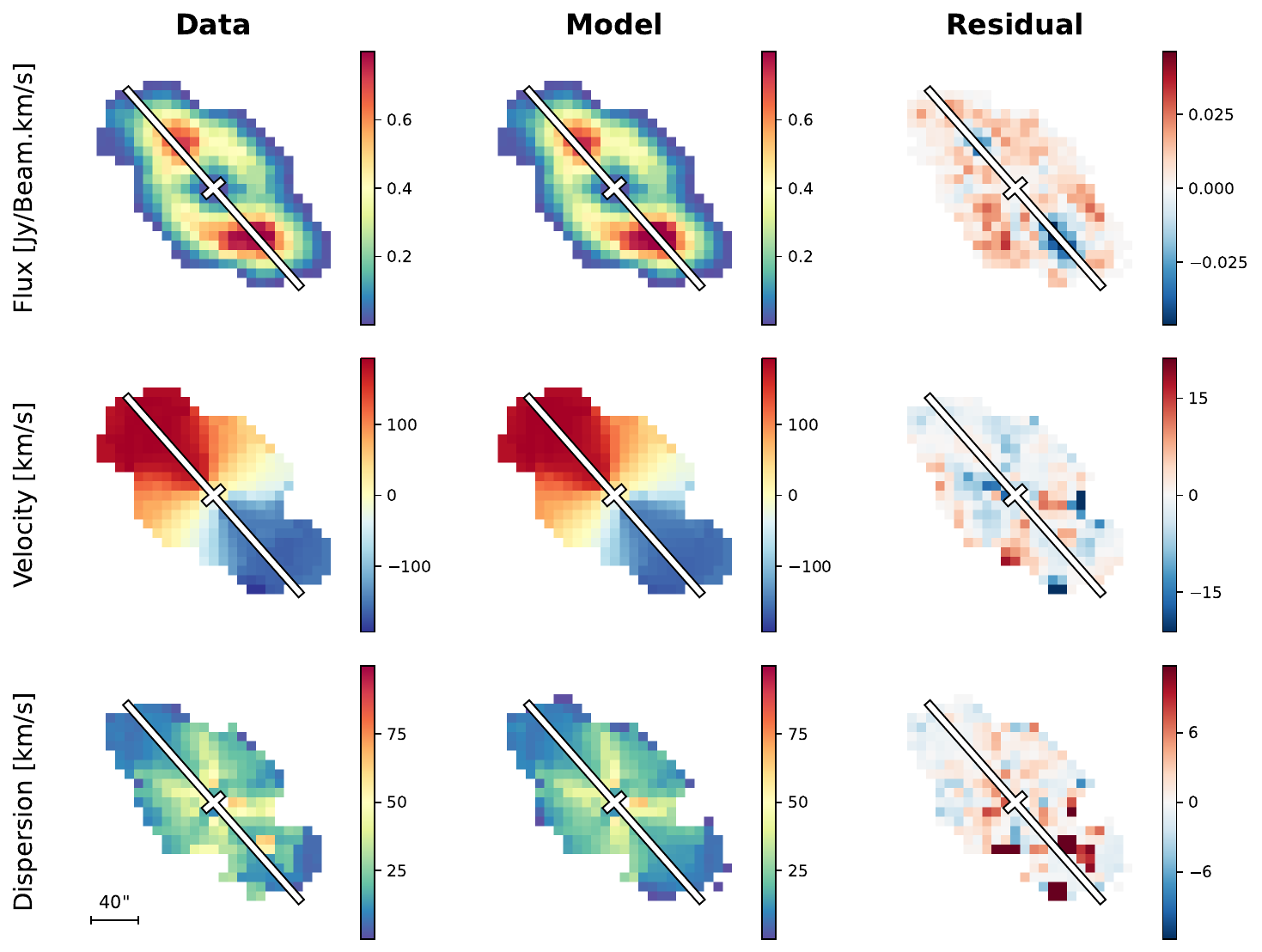}
    \caption{\lrto{The observed and modeled 2D maps for UGC 4503.} The rows from top to bottom display \HI flux maps, velocity maps and velocity dispersion maps respectively. The columns represent the observed GMRT data, the best-fitting MCMC model, and the residuals. The long white solid line indicate the position of the kinematic major axis. The color bars indicate the scale for each parameter. 
    }
    \label{fig:dysmalpy_2d}
\end{figure*}

\begin{figure*}
    \centering
    \includegraphics[width=0.9\linewidth]{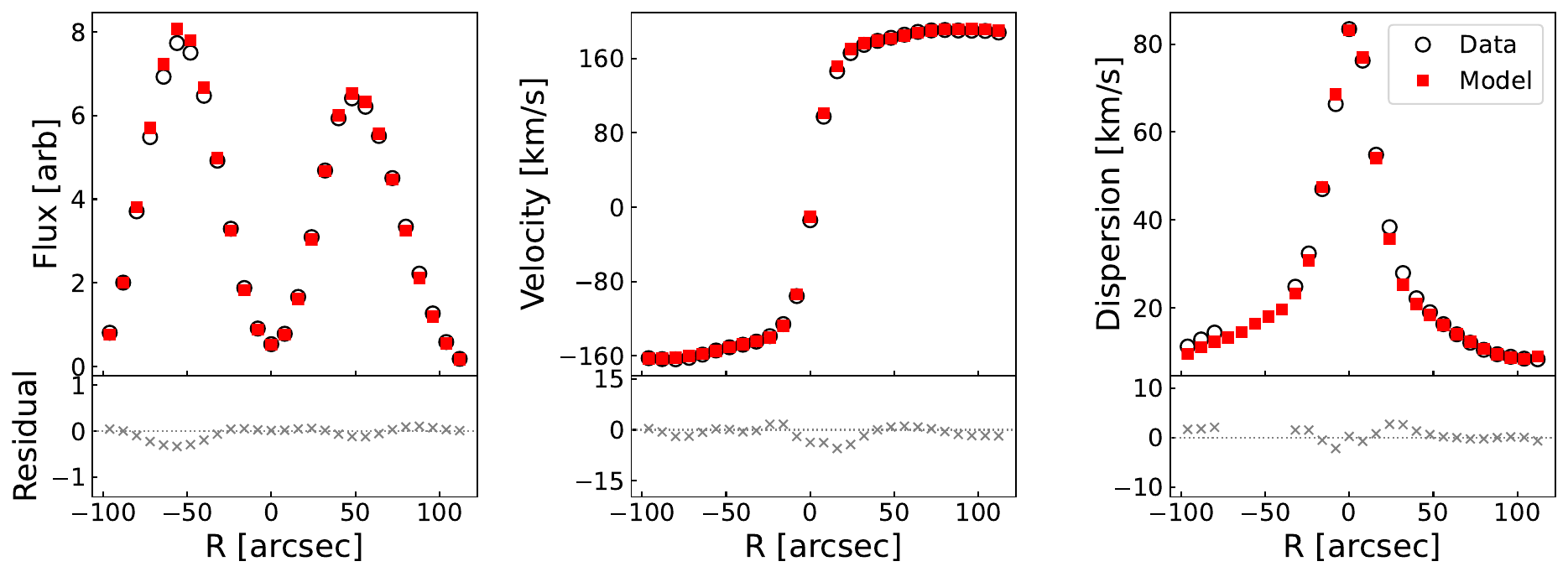}
    \caption{\lrto{The observed and modeled 1D extractions for UGC 4503.} The 1D radial extraction is along the major axis (indicated by the white slit in Figure \ref{fig:dysmalpy_2d}). The top plots show the comparison between the observed data (black circles) and the best-fitting model (red squares) for flux, velocity and velocity dispersion. The bottom plots display the corresponding residuals.}
    \label{fig:dysmalpy_1d}
\end{figure*}

Figure \ref{fig:dysmalpy_2d} presents the spatially resolved kinematic maps comparing the observational data with our best-fitting dynamical model. The model reproduces the overall morphology in moment maps. The residual map for the velocity field is largely featureless with a root-mean-square residual of $V_{\rm rms}$=11.6 km/s, confirming the robustness of the rotation curve fit.
To quantitatively assess the quality of the fit, we extracted 1D profiles along the major axis, as shown in Figure \ref{fig:dysmalpy_1d}. The model of the velocity field shows agreement with the data, tracing the steep rise in the inner region and the flattening of the curve at larger radii ($\left| r \right|>20$ arcsec).
The velocity dispersion profile exhibits a central peak 
and decreases towards the outer radii. Our model generally follows the global observed trend.

\begin{figure*}
    \centering
    \includegraphics[width=0.9\linewidth]{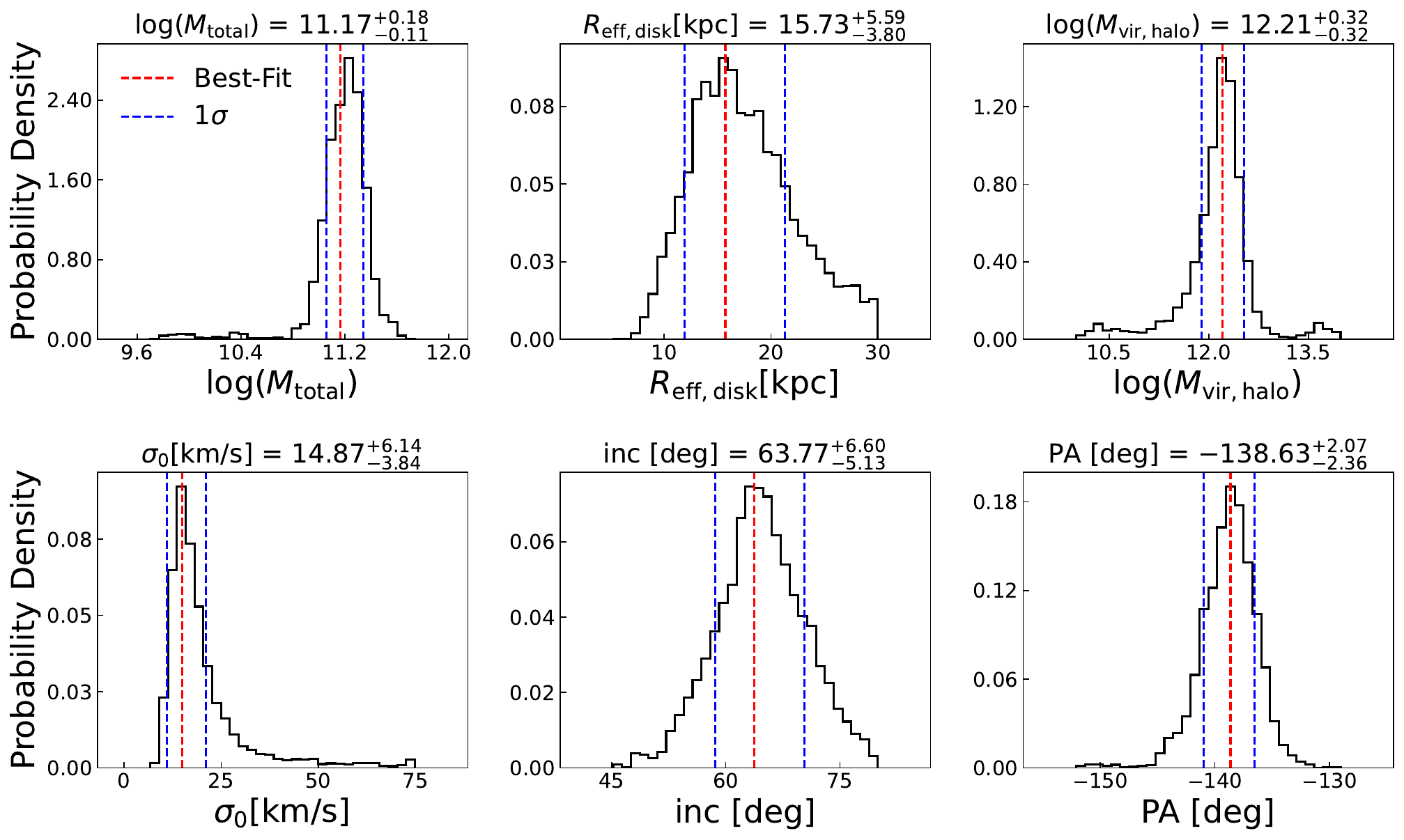}
    \caption{1D posterior probability distributions of UGC 4503 from our \textsc{DysmalPy} MCMC fitting. From left to right and top to bottom are total mass in logarithm solar unit ($\log(M_{\rm total})$), effective radius of the disk ($R_{\rm eff, disk} [\rm {kpc}]$), logarithm of dark matter halo virial mass ($\log(M_{\rm vir, halo})$), intrinsic velocity dispersion ($\sigma_0 [\rm{km/s}]$), inclination and position angle. The red and blue vertical dashed lines mark the best-fit and 1-$\sigma$ uncertainty. Best-fit values with errors are also shown above the graphs.
   }
    \label{fig:1d_histograms}
\end{figure*}

From the 1D \lrto{histogram} plots of unfixed parameters displayed in Figure \ref{fig:1d_histograms}, it is evident that all parameters exhibit clear \lrtM{single-peak post-probability distributions,} indicating that the MCMC sampling has reached convergence. The total mass, halo virial mass, and the geometric parameters (inclination and position angle) are tightly constrained by the observational data. The effective disk radius and velocity dispersion show slightly broader distributions and the whole trend is well-constrained with the uncertainties in a reasonable range. Overall, the fitting results are robust and reliable. \lrtnew{The best-fit value of the intrinsic velocity dispersion is \lrtM{$14.87^{+6.14}_{-3.84} \rm {km/s}$}, which has been decoupled from instrumental smearing effects via {\sc DysmalPy} 3D forward modeling with both point spread function (PSF) and line spread function (LSF) convolutions}. This value is higher than the characteristic value of the \HI velocity dispersion of 10 $\pm$ 2 km/s \citep{tamburro2009driving}.

\begin{figure*}
    \centering
    \includegraphics[width=0.7\linewidth]{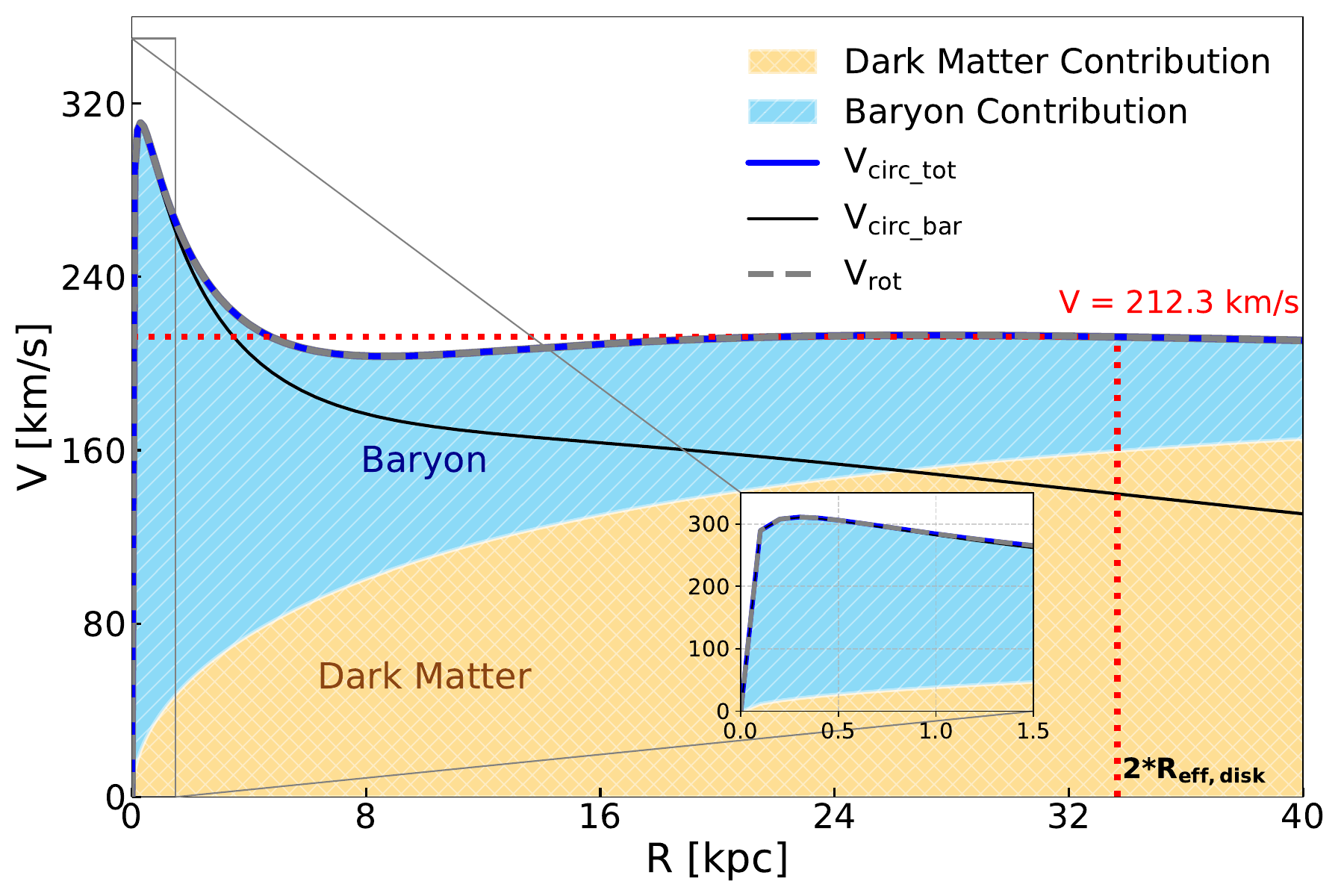}
    \caption{\lrtM{Intrinsic} rotation curve of UGC 4503. The inset provides a zoom-in view of the rise in rotational curve within 1.5 kpc. The total circular velocity ($V_{\rm {circ\_tot}}$) profile is shown as thick blue lines. The black line indicates baryonic matter velocity ($V_{\rm {circ\_bar}}$) and the gray dashed line  means the rotational velocity ($V_{\rm rot}$) considering asymmetric drift correction. The baryonic and dark matter contributions to the circular velocity are plotted in light blue and orange shading respectively and stacked on each other. The dashed vertical gray line displays the effective disk radius ($R_{\mathrm{eff,\,disk}}$) of UGC 4503. The \lrto{red} marks locate the rotational velocity at $2\times R_{\mathrm{eff,\,disk}}$.
    }
    \label{fig:rotationcurve}
\end{figure*}

Based on the above modeling in \textsc{DysmalPy}, we obtain the intrinsic rotation curve of UGC 4503 in Figure \ref{fig:rotationcurve}. Figure \ref{fig:rotationcurve} presents the radial profile of the galaxy's rotation curve. The total circular velocity (thick blue line) exhibits a central rise, peaking within the inner 1-2 kpc (as detailed in the inset). This feature indicates a highly concentrated central mass distribution. Following the central peak, the velocity profile declines slightly. After reaching the effective disk radius, it remains stable. 

The rotational velocity at $2\times R_{\mathrm{eff,\,disk}}$ is approximately 212.3 km/s and combining the fitted result of intrinsic velocity dispersion in Figure \ref{fig:1d_histograms} (\lrtM{$\sigma_0=14.87^{+6.14}_{-3.84}$ km/s}), we calculate the $V/\sigma$ of UGC 4503 is \lrtM{$14.28_{-4.17}^{+4.97}$}, which is lower than the typical values ($\approx20$) reported in local spiral galaxies \citep{battaglia2006study}. The suppressed $V/\sigma$ may stem from the elevated velocity dispersion. This could imply the influence of AGN feedback on disk formation. The central nucleus releases substantial amounts of energy through jets and/or outflows. The energetic flows interact with the ISM and induce turbulence, \lrtM{and materials falling back onto disk planes may also contribute to the velocity dispersion in the disk.} \lrt{Several previous investigations have also pointed out that various physical processes, most notably stellar winds and AGN activity, have been proposed as energy sources capable of driving \HI turbulence and elevating \HI velocity dispersion \citep{das2020tracing, tamburro2009driving, krumholz2016turbulence,Yu_2022},} \lrtM{although a more detailed understanding still requires a much higher angular resolution kinematics data.}

\section{Future Survey Expansion}
\label{sec:future}

\lrtP{In addition to our results of the \HI mass-size relation and the kinematic signature of AGN feedback in a case study of UGC 4503, AGN feedback has been studied by both observations and simulations.}

\lrtP{From \HI observations,} a stacking result for measuring the average \HI content of AGN host galaxies ($M_{*} > 10^{10} \rm M{_\odot}$) indicates that AGNs do not influence the large-scale gaseous properties of galaxies in the local Universe \citep{fabello2011alfalfa}. The \HI gas fraction shows no significant difference between the non-AGN and AGN samples at all values of black hole accretion rate. \lrtp{Similarly, \cite{ellison2019atomic} suggests that global depletion of gas reservoirs via AGN feedback is not a dominant mechanism in low-redshift galaxies.} \lrtP{While these observational results appear to be at odds with theoretical expectations of AGN feedback effects, this discrepancy may be attributed to black holes primarily impacting the central regions, while the \HI gas is detected on much larger scales.}

\lrtP{On the other hand,} phenomena of AGN feedback are seen in the ionized gas phase from Mapping Nearby Galaxies at Apache Point Observatory (MaNGA) observations \citep{bundy2014overview}. Highly elevated velocity dispersion in the center is boosted by AGN activities \citep{deconto-machadoIonisedGasKinematics2022}, ionized gas outflows are significantly more prevalent in AGN hosts compared to those in non-AGNs \citep{Wylezalek2020manga} and mass outflow rate correlates with AGN luminosity showing the role of AGN in driving galactic-scale winds \citep{Avery2021manga}. 

Moreover, AGNs have been found to suppress central star formation rates \citep{Lammers2023manga} and AGN feedback in quenched low-mass galaxies \lrtP{also} prevents gas cooling \citep{Penny2018manga}. 

\lrtP{As for metallicities, AGN can halt accretion and drive galaxies off star-forming main sequence, causing metallicity to remain constant while star formation rates decline \citep{Li2024metallicity} and AGN hosts have smaller correlation lengths ($l_{\rm corr}$) at fixed mass but similar $l_{\rm corr}$ at fixed star formation rates (SFRs), suggesting AGN suppress SFRs and thus $l_{\rm corr}$ \citep{Li2025metallicitycorrelations}}. Besides, AGN narrow line regions often exhibit lower metallicities compared to those of the host galaxy disk, suggesting that AGN activity can be fueled by the accretion of metal-poor external gas \citep{Nascimento2022manga}. 

Furthermore, current hydrodynamic simulation models still struggle to match the detailed attributes of \HI gas \citep{wang2024hireview}.
\lrtp{Recent TNG simulations over-estimate the \HI disc size at a given \HI mass and over-predict central deficit in \HI \citep{Diemer2019tng,Gebek2023tng,stevens2023tng}. 
The \HI scaling relations also provide evidence that the AGN feedback model in TNG simulations requires refinement \citep{wang2024hireview}. In TNG100, star-formation-quenched galaxies retain excess \HI for their stellar mass, as AGN feedback redistributes rather than depletes the cold gas \citep{ma2022tng}. Therefore, the properties of \HI gas remain the bottleneck in achieving a comprehensive picture of AGN feedback. We will obtain the \HI gas maps of AGN-host galaxies in a larger sample to perform a systematic comparison of the multi-phase outflow properties.}

\lrtp{The FAST all sky \HI (FASHI) survey \citep{zhangFASTAllSky2023} is conducted using the Five-hundred-meter Aperture Spherical radio Telescope (FAST) covering a total of 41741 extragalactic \HI sources to a redshift limit of $z \lesssim 0.09$ and has a median detection sensitivity of around 0.76 mJy/beam with a spectral line velocity resolution of $\sim$ 6.4 km/s at a frequency of $\sim$ 1.4 GHz.} 
\lrtP{By cross-matching the FASHI and SDSS DR7 catalogs, we selected a sample of Seyfert galaxies with significant \HI content and well-defined rotational kinematics, which provide the necessary constraints for investigating AGN-gas interactions.} From MaNGA observations, we notice that all the target Seyfert galaxies have potential signatures of AGN-driven outflows. The [O III]$\lambda$5007 velocity map shows orderly rotation, consistent with the \HI velocity diagram shown in FASHI, whereas the elevation of central velocity dispersion and the strong Na ID absorption indicate strong gas outflows powered by the active nucleus.

A particularly promising development of our on-going work is the FAST Core Array \citep{jiangFASTCoreArray2024,JiangPeng2025}, which is designed to combine the unparalleled sensitivity of FAST with high angular resolution (4.3$''$ at 1.4 GHz). 
There will be two phases for the construction of the FAST Core Array.
In the first phase, there will be 24 antennas with 40 m diameters, while this number will increase to 64 in the second phase. 
These antennas are distributed within a 5 km radius around FAST, enabling interferometric imaging at resolutions previously inaccessible to single-dish observations. 
By coupling FAST’s exceptional sensitivity with the angular resolution of the core interferometer and leveraging state-of-the-art instrumentation, the FAST Core Array will dramatically expand the sample of AGN host galaxies with spatially resolved \HI mapping. Resolving the atomic gas structure in galaxies is a primary scientific objective of the array, making it ideally suited for systematic studies of AGN feedback in the \HI phase.

\lrtP{In the future, we will constrain the impact of AGN activity on the multi-phase gas cycle, shedding light on its role in driving galaxy evolution across different environments and evolutionary stages.}

\section{Summary}
\label{sec:sum}

In this paper, we describe the first systematic, high-resolution \HI survey of 8 local Seyfert galaxies using GMRT observations, probing the impact of AGN feedback on the distribution and kinematics of \HI gas reservoirs. We utilize CASA and the GMRT-pipeline in the data reduction and CARTA for the visualization of \HI spectra and moment maps. \lrtnew{Subsequently, we derive their global atomic gas properties, including \HI masses and sizes and display two representative kinematic maps for two galaxies (UGC 4503 and UGC 9535).}

In the \HI mass-size relation, emcee method is used for linear regression. 
\lrtb{We find that AGN-host galaxies exhibit a slightly shallower \hi mass–size relation than the canonical relation; however, the measured slope remains consistent with the canonical value within $2\sigma$ uncertainties}, which imply that AGN feedback do not seem to significantly influence \HI distribution in the outer disks \lrto{(see Figure \ref{fig:mass_size}). }
 
We highlight the dynamical modeling of UGC 4503 using {\sc DysmalPy} software to investigate the link between AGN feedback and disk dynamics. Employing \textsc{DysmalPy}, we achieve reconstruction of the \HI kinematics: 2D moment maps and 1D extraction profiles. 
The MCMC analysis shows well-constrained unimodal posterior distributions for key parameters. The intrinsic rotation curve of UGC 4503 is also presented. From MCMC fitting, we derive an intrinsic velocity dispersion of \lrtM{$14.87^{+6.14}_{-3.84}$ km/s}, higher than the typical \HI dispersion of $\sim$ 10 km/s. Consequently, the calculated $V/\sigma$ ratio is \lrtM{$14.28_{-4.17}^{+4.97}$}, which is lower than that of typical local spiral galaxies. The suppressed $V/\sigma$ ratio might provide evidence for AGN feedback on the host galaxy, where outflows and/or jets from the central nucleus prevent the formation of a cold disk.

In future research, we will expand our Seyfert galaxy sample. Our future work will focus on the joint observations and analysis using GMRT, FAST and MaNGA. GMRT can provide the spatially resolved data of cool gas and we can get detailed ionized gas velocity fields, star formation rates, and metallicity gradients from MaNGA IFU. Our survey will further research \HI kinematics for signatures of AGN-driven outflows and quantify feedback effects on star formation efficiency through spatially resolved analysis of \HI depletion timescales, thereby advancing a unified picture of multi-phase AGN feedback in galaxy evolution.

\begin{acknowledgments}
\section*{acknowledgments} 
The authors express their gratitude to the anonymous reviewer for constructive comments, which significantly improved the manuscript. We thank Peng Jiang for useful discussion focused on the FAST Core Array.
This work is supported by the National Key R\&D Program of China No.2025YFF0510603, the National Natural Science Foundation of China (grant 12373009), the CAS Project for Young Scientists in Basic Research Grant No. YSBR-062, the China Manned Space Program with grant no. CMS-CSST-2025-A06, and the Fundamental Research Funds for the Central Universities. XW acknowledges the support by the Xiaomi Young Talents Program, and the work carried out, in part, at the Swinburne University of Technology, sponsored by the ACAMAR visiting fellowship.

We thank the staff of the GMRT that made these observations possible. GMRT is run by the National Centre for Radio Astrophysics of the Tata Institute of Fundamental Research.

\end{acknowledgments}

\software{Astropy \citep{astropy:2022}, CASA \citep{mcmullin2007casa}, CARTA \citep{ott2020carta}, {\sc Dysmal}/{\sc DysmalPy} \citep{2004ApJ...602..148D,2004ApJ...613..781D,2009ApJ...697..115C,2011ApJ...741...69D,2016ApJ...831..149W,2017ApJ...840...92L,2021ApJ...922..143P}}
\vspace{5mm}

\bibliography{msref}{}
\bibliographystyle{aasjournal}

\end{document}